\documentclass[aps,pra,reprint,twocolumn,superscriptaddress,longbibliography,floatfix,citeautoscript]{revtex4-2}
\usepackage[normalem]{ulem}
\usepackage{graphicx} 
\usepackage{hyperref}

\usepackage{tikz}
\usetikzlibrary{quantikz2}
\usepackage{amsmath,amsfonts,amssymb,bm}
\usepackage{tikz}
\usepackage[left=1.5cm, right=1.5cm, top=2.5cm, bottom=2.5cm]{geometry}

\usetikzlibrary{arrows.meta,decorations.pathreplacing}

\newcommand{\lc}[1]{{#1}}

\newcommand{\as}[1]{#1}
\DeclareRobustCommand{\asdeleted}[1]{\ifmmode\mathpalette\asdeletedmath{#1}\else\sout{#1}\fi}
\newcommand{\asdeletedmath}[2]{\text{\sout{\ensuremath{#1#2}}}}
\newcommand{\affOxfordPTCL}{\affiliation{Physical and Theoretical Chemistry Laboratory, Department of Chemistry, University of Oxford, Oxford OX1 3QZ, UK}}
\newcommand{\affOxfordPhys}{\affiliation{Department of Physics, Clarendon Laboratory, University of Oxford, Oxford OX1 3QU, United Kingdom}}

\newcommand{\affAnB}{\affiliation{Alice \& Bob, 53 Bd du Général Martial Valin, 75015 Paris, France}}
\newcommand{\affNPL}{\affiliation{National Physical Laboratory, Teddington, TW11 0LW, United Kingdom}}

\begin{document}
\title{Exploiting SU(2) Symmetry for Efficient Quantum State Preparation of Matrix Product States}

\author{Etienne D'Argenlieu}
\affOxfordPhys  
\author{Alessandro Summer}
\affAnB
\author{Fran\c{c}ois Jamet}
\affAnB
\author{David P. Tew}
\affOxfordPTCL
\author{Lachlan P. Lindoy}
\email{lachlan.lindoy@npl.co.uk}
\affNPL

\begin{abstract}
Quantum computing offers a promising route to simulating strongly correlated systems, particularly for dynamical and spectral quantities whose classical computation can become prohibitively expensive. 
Fault-tolerant algorithms for these tasks, including real-time dynamics and phase estimation, assume efficient preparation of the initial many-body state, making state preparation an important bottleneck. Matrix product states obtained classically with the density matrix renormalization group can be compiled sequentially into fault-tolerant circuits. 
We exploit non-Abelian spin SU(2) symmetry in this compilation. By the Wigner-Eckart theorem, only the reduced tensors, which carry the variational information, need generic unitary synthesis. The magnetic structure is restored by Clebsch--Gordan transformations with analytically known angles. We benchmark the method on representative strongly correlated models, including the one-dimensional Hubbard model, quasi-one-dimensional Hubbard geometries, and Anderson impurity models, and perform fault-tolerant resource estimates using the Toffoli count as the primary cost metric. Across the systems considered, exploiting non-Abelian SU(2) symmetry reduces the Toffoli cost roughly threefold relative to U(1)-adapted compilation and by more than an order of magnitude relative to the standard dense compilation method. These results demonstrate how non-Abelian symmetries can substantially reduce the cost of transferring classically tractable many-body ground states to quantum computers, facilitating their use in quantum dynamics and energy estimation. 
\end{abstract}

\maketitle
\section{INTRODUCTION}

The simulation of strongly correlated quantum systems remains a major computational challenge in condensed-matter physics and quantum chemistry. The dimension of the many-body Hilbert space grows exponentially with system size, severely limiting direct classical approaches. Tensor-network methods, and in particular the density matrix renormalization group (DMRG) \cite{White1992,white1993,Schollwock2005, Roman}, circumvent this exponential complexity for an important class of systems by representing low-entanglement states compactly as matrix product states (MPS) \cite{Schollwock2011}. This has made DMRG a powerful tool for computing ground-state properties of one-dimensional and quasi-one-dimensional quantum systems \cite{Verstraete2006, Hastings2007, Eisert2010}.

\begin{figure*}[t]
  \centering
  \includegraphics[width=\textwidth]{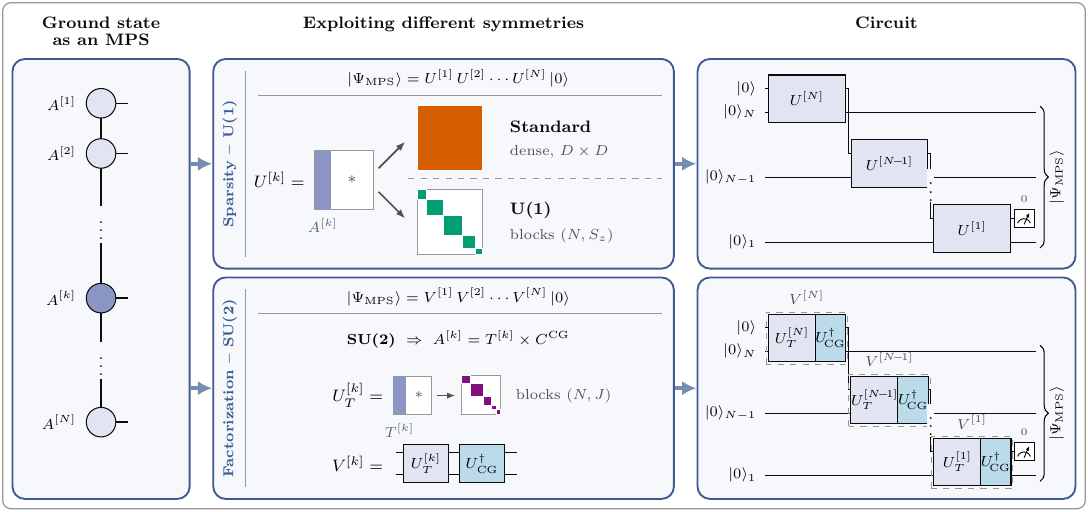}
  \caption{\textbf{Overview of symmetry-adapted MPS state preparation.}
Left: the ground state is represented as a matrix product state (MPS) with site tensors $A^{[k]}$. In canonical form, each $A^{[k]}$ is an isometry.
Middle: each isometry is embedded in the first columns (shaded) of a unitary $U^{[k]}$, the remaining columns ($\ast$) being a free completion, so that $|\Psi_{\mathrm{MPS}}\rangle = U^{[1]}U^{[2]}\cdots U^{[N]}|0\rangle$. Without symmetry, $U^{[k]}$ is dense (orange); with U(1) symmetry, it is block diagonal in $(N,S_z)$ sectors (green). With SU(2) symmetry, the Wigner--Eckart theorem factorizes each tensor as $A^{[k]}=T^{[k]}\times C^{\mathrm{CG}}$, where $T^{[k]}$ is a reduced tensor and $C^{\mathrm{CG}}$ contains Clebsch--Gordan coefficients. Only $T^{[k]}$ is embedded in a unitary $U_{T}^{[k]}$, which is block diagonal in $(N,J)$ sectors (purple) and acts on the multiplicity space only, with $J$ the total spin of the bond. The magnetic structure is then restored analytically by an inverse Clebsch--Gordan transformation $U^{\dagger}_{\mathrm{CG}}$ (light blue), applied as a SELECT operation conditioned on the coupling channel, giving $V^{[k]}=\mathrm{SELECT}(U^{\dagger}_{\mathrm{CG}})\,U_{T}^{[k]}$ and $|\Psi_{\mathrm{MPS}}\rangle = V^{[1]}V^{[2]}\cdots V^{[N]}|0\rangle$.
Right: the corresponding sequential state-preparation circuits. Starting from site $N$, the bond register, initialised in $|0\rangle$, interacts successively with each physical register $|0\rangle_k$ through $U^{[k]}$ (top) or $V^{[k]}$ (bottom). After the last step, the bond register returns to $|0\rangle$, as verified by the final measurement, leaving the physical registers in $|\Psi_{\mathrm{MPS}}\rangle$. Matrix and block sizes are schematic.}
  \label{fig:overview}
  \label{fig:MPStoUnitaries}
\label{fig:SU2QC}
\end{figure*}

The situation becomes more challenging when dynamical properties are considered. Real-time evolution can generate entanglement rapidly, causing the bond dimension required for an accurate MPS representation to grow and
eventually making classical tensor-network simulations prohibitively
expensive \cite{schuch2008entropy}. 
Quantum computers provide a natural alternative for such calculations, since quantum states and their unitary dynamics are represented directly in a quantum Hilbert space \cite{Lloyd1996,Georgescu2014,McArdle2020}. This
suggests a hybrid classical--quantum strategy in which a correlated ground state is first obtained classically using DMRG, transferred to a quantum computer, and subsequently used as the initial state for quantum simulations of dynamical quantities \cite{Jamet2025}. 

Such a strategy is particularly relevant to quantum
impurity problems, including those arising in dynamical mean-field theory (DMFT) \cite{Georges1996}. In DMFT, the
impurity problem is solved repeatedly until self-consistency is reached, and
each iteration requires the ground state of a new impurity Hamiltonian. Within an iteration, the impurity Green's function
is obtained from circuits that prepare this ground state, evolve it in time and
measure it. Since the measurement destroys the state, the ground state has to be
prepared again for every time point, every component of the Green's function and
every repetition of the measurement. State preparation is therefore repeated a
large number of times in a single calculation. State preparation is also an important step in quantum phase estimation, whose success probability is set by the overlap
between the initial state and the target eigenstate. A good initial state is
therefore essential, as discussed in general for quantum
chemistry~\cite{Fomichev2024} and for the Hubbard model~\cite{Yoshioka2024}. In this work, we follow the approach of  Ref.~\cite{berry2025rapid, schon2005sequential} and prepare the initial state from an MPS.

An essential requirement of this approach is the efficient preparation of the classically obtained MPS on a quantum computer. 
An MPS in canonical form can be interpreted as a sequence of isometries and
prepared sequentially by embedding these isometries into quantum
circuits~\cite{schon2005sequential,berry2025rapid}. It can also be approximated
by shallower circuits, built from layers of two-qubit
gates~\cite{PhysRevA.101.032310,Rudolph_2024}, optimised
variationally~\cite{PRXQuantum.2.010342,Dborin_2022,10.1145/3731251,m5kb-4f43},
or of logarithmic depth~\cite{PhysRevLett.132.040404,Szołdra_2026}, and such
preparations have been run on noisy quantum
hardware~\cite{PhysRevA.109.062437,Pennington2026}. However, variational approaches are subject to optimisation overheads that are difficult to predict.  As such, here we focus on exact
sequential preparation. Although it avoids preparing an arbitrary exponentially
large state vector, its circuit cost (as measured in terms of Toffoli count) can remain substantial for MPS with large
bond dimension. In a fault-tolerant setting this cost matters, because
non-Clifford gates are typically implemented through magic-state
distillation~\cite{bravyi2005universal,Litinski2019}. Reducing the non-Clifford
cost of MPS state preparation is therefore an important step toward using
classically generated many-body states as inputs to fault-tolerant quantum
algorithms.

A natural route to reducing this cost is to exploit physical symmetries already
encoded in the tensor-network representation. Many electronic Hamiltonians conserve the particle number and the spin
projection $S_z$ \cite{Szabo1996}, giving rise to global U(1) symmetries. Incorporating
this symmetry into an MPS imposes selection rules on its tensors and decomposes
them into symmetry-compatible blocks \cite{McCulloch2007,Singh2010,Singh2011U1}. Recent work
has shown that this block structure can be exploited directly during quantum
circuit synthesis, substantially reducing the resources required for MPS state
preparation compared with symmetry-agnostic compilation
\cite{rupprecht2026faster}. However, many physically relevant Hamiltonians possess additional non-Abelian symmetries whose structure is not captured by U(1) symmetry alone.

In particular, spin-rotation-invariant electronic Hamiltonians possess an
SU(2) spin symmetry  \cite{Szabo1996}. Unlike the Abelian U(1) case, SU(2) symmetry not
only restricts which tensor blocks are allowed, but also fixes the dependence
of the tensors on the magnetic quantum numbers. An SU(2)-symmetric MPS can
be decomposed, through the Wigner--Eckart theorem, into reduced tensors
containing the independent variational degrees of freedom and structural
tensors determined entirely by Clebsch--Gordan coefficients
\cite{McCulloch2007,Singh2012SU2,WEICHSELBAUM2012,Schmoll2019SU2}. Consequently, explicitly expanding an SU(2)-adapted MPS into all of its
magnetic components before quantum compilation introduces degrees of freedom
that are fixed by symmetry rather than independently specified by the many-body
state. Symmetries have already been used in other approaches for quantum state
preparation. Approximate initial states can be filtered by projectors onto the
target particle-number and total-spin sectors before quantum phase
estimation~\cite{Khinevich2026}, and spin eigenfunctions can be prepared directly
by circuits that implement the addition of angular momenta~\cite{Carbone2022,PhysRevResearch.7.013191}.
These approaches act on the full many-body state. Here, instead, the SU(2)
symmetry is exploited locally, in the compilation of each MPS tensor, so that
only the degrees of freedom that are not fixed by symmetry are synthesized.

In this work, we exploit this observation to develop an
SU(2)-adapted quantum state-preparation method for MPS. Building on existing
sequential MPS preparation and U(1)-adapted compilation techniques
\cite{berry2025rapid,rupprecht2026faster}, we compile the reduced MPS tensors
directly and reconstruct the magnetic degrees of freedom using inverse
Clebsch--Gordan transformations. The resulting procedure avoids explicitly
synthesizing the full spin-resolved tensors and reduces the dimension of the
variational spaces that must be compiled. We benchmark the approach on Hubbard chains, Hubbard cylinders
and multi-orbital Anderson impurity models of SrVO$_3$, using the Toffoli
count as the fault-tolerant cost metric. At equal angle precision, our
analytical estimates give U(1)-to-SU(2) and dense-to-SU(2) Toffoli-count
ratios of $2.4$--$3.7$ and $26$--$68$, respectively (Fig.~\ref{fig:overview}), for
the states in Table \ref{tab:toffoli-all} at $b = 21$. The estimated gain
grows with bond dimension in the truncation scans.

\lc{ While this manuscript was being finalised, Hennefarth \textit{et al.}~\cite{hennefarth2026quantumloadingchargespinadapted} independently presented an SU(2)-adapted MPS preparation scheme based on Householder-LKS synthesis, along with resource estimates for Fermi-Hubbard lattices and molecular systems, including iron-sulfur clusters, with comparisons to dense MPS preparation. They also outline direct synthesis as an alternative, without implementing or costing it. Here we use Clements synthesis and focus on the gain over U(1)-adapted compilation using the same synthesis method, additionally we identify how angle precision in the preparation impacts preparation accuracy for the different schemes.}

\section{MPS STATE PREPARATION}
\label{sec:mps_prep}

A quantum system composed of $N$ sites, each with local Hilbert-space dimension $d$, is described by a state in a Hilbert space of dimension $d^N$. A matrix product state (MPS) provides a compact representation of such a many-body state by decomposing its coefficient tensor into a product of local tensors,
\begin{equation}
|\Psi\rangle =
\sum_{i_1,\ldots,i_N=1}^{d}
A^{[1]i_1}
A^{[2]i_2}
\cdots
A^{[N]i_N}
|i_1 i_2 \cdots i_N\rangle,
\end{equation}
where $A^{[k]i_k} \in \mathbb{C}^{\chi_{k-1}\times\chi_k}$ is the local tensor associated with site $k$, and $\chi_k$ denotes the corresponding bond dimension, with $\chi_0=\chi_N=1$ for open boundary conditions \cite{Schollwock2011,Roman}. We only need $O(Nd\chi^{2})$ coefficients to describe the state. 

In left-canonical form, the MPS tensors are isometries and can therefore be extended to unitary operators. Ref.\cite{schon2005sequential} provides a method for constructing a quantum circuit from these unitaries, as illustrated in Fig.~\ref{fig:MPStoUnitaries}. The problem now becomes a problem of unitary synthesis, and it has been studied in several works \cite{Low2024LKS, berry2025rapid, rupprecht2026faster}. We build on the general method developed in Ref.\cite{berry2025rapid} for standard MPS, and extended recently to Abelian symmetries in Ref.\cite{rupprecht2026faster}.

Following Ref.~\cite{berry2025rapid}, the generic
unitaries appearing in the sequential generation circuit (Fig.~\ref{fig:MPStoUnitaries}) are synthesized using the
Clements decomposition \cite{Clements2016}, which expresses an arbitrary $N\times N$
unitary as a rectangular mesh of two-level transformations acting on
pairs of neighboring modes (Appendix~\ref{app:compil}).

\as{Left-canonical tensors satisfy $\sum_{i_k}(A^{[k]i_k})^\dagger A^{[k]i_k}=I_{\chi_k}$. At site $k$, the isometry takes bond $k$ to bond $k-1$ and emits site $k$. Writing $U^{[k]}$ for its unitary extension, preparation starts at site $N$, giving}
\begin{equation}
|\Psi_{\mathrm{MPS}}\rangle
=
\as{{U^{[1]}U^{[2]}\cdots U^{[N]}}}
|\Psi_{\mathrm{init}}\rangle ,
\end{equation}
where $|\Psi_{\mathrm{init}}\rangle$ denotes the register initialised to zero.
The total preparation cost is therefore determined by the cost of
synthesizing the local MPS isometries. For approximately uniform bond
dimension $\chi$ and $b$ bits of precision for the programmable
rotations, the construction of Ref.~\cite{berry2025rapid} achieves a
Toffoli complexity scaling as

\begin{equation}
\begin{aligned}
C_{\rm tot}={}& \as{{f(D_N)}}+\sum_{i}\Big[
  \Big\lceil\tfrac{D_i}{\Lambda}\Big\rceil+\Lambda b-2 \\
&\quad+D_i\Big(\Big\lceil\tfrac{D_i}{\Lambda}\Big\rceil+2\Lambda b-5\Big)
  +(n-3)(D_i-1) \\
&\quad+\Big\lceil\tfrac{D_i}{\Lambda}\Big\rceil+2\Lambda b \\
&\quad+\Big\lceil\tfrac{2D_i}{\Lambda'}\Big\rceil+\Lambda'-6
\Big] .
\end{aligned}
\label{eq:total-cost}
\end{equation}
Here $D_i=\max(\chi_{i-1},\chi_i)$, \as{ the register has $n=\lceil\log_2(2D_i)\rceil$ qubits,} $\Lambda,\Lambda'$ are the space--time trade-off
parameters of the QROM, optimised per site, and $\as{{f(D_N)}}$ is the cost of
the first unitary (Appendix~\ref{app:complexity}). For a local dimension
$d>2$ the cost is multiplied by $d-1$~\cite{berry2025rapid}, i.e.\ by $3$
for a spatial orbital.

When the MPS carries a U(1) symmetry, the tensors decompose into
blocks labelled by the conserved charge. Ref.~\cite{rupprecht2026faster}
exploits this structure directly at the compilation level. Rather than
synthesizing a dense unitary over the complete bond space, the
symmetry-compatible blocks are synthesized independently. If the
corresponding block dimensions are $D_q$, the Clements decomposition
is applied separately within each symmetry sector,
\begin{equation}
    U^{[k]} = \bigoplus_q U_q^{[k]}.
    \label{eq:u1_blocks}
\end{equation}
The synthesis is therefore governed by the dimensions of the
symmetry-compatible blocks rather than by the dimension of the full
bond space. In particular, the blocks are synthesised in parallel, so that the number
of Clements layers is set by the largest block,
\begin{equation}
  D_{\max}=\max_q D_q ,
\end{equation}
while the QROM of each layer still loads the angles of all \as{active} blocks. For
molecular MPS this reduces the Toffoli count by a factor of $10$--$30$
compared with the dense compilation~\cite{rupprecht2026faster}.

The key observation underlying the present work is that a non-Abelian
SU(2) symmetry provides additional structure beyond this block
sparsity. In particular, the magnetic components within a spin
multiplet are not independent variational degrees of freedom, but are
fixed by SU(2) representation theory. In the following section, we
show how this structure can be exploited to compile only the reduced
MPS tensors and reconstruct the magnetic degrees of freedom through
Clebsch--Gordan transformations.

\section{SU(2)-adapted synthesis}
\label{sec:su2_method}

We now introduce an SU(2)-adapted synthesis procedure that exploits
the non-Abelian spin structure of the MPS. The central idea is to avoid reconstructing and
synthesizing the complete spin-resolved MPS tensors. Instead, we compile
only their reduced, variational part and reconstruct the magnetic
degrees of freedom using Clebsch--Gordan (CG) transformations (Fig.~\ref{fig:MPStoUnitaries}).

\subsection{SU(2)-adapted MPS representation}

For a spin-rotation-invariant MPS that also conserves the particle
number, the Hilbert space on bond $k$ decomposes into irreducible
representations of $U(1)\times SU(2)$,
\begin{equation}
    \mathcal{H}_k
    =
    \bigoplus_{q=(N,J)}
    \left(
        \mathcal{M}_{k,q}
        \otimes
        \mathcal{V}_{J}
    \right),
    \label{eq:su2_bond_decomposition}
\end{equation}
where $\mathcal{V}_{J}$ is the spin-$J$ irreducible representation,
of dimension $2J+1$, and $\mathcal{M}_{k,q}$ is the multiplicity
space of the symmetry sector $q=(N,J)$, of dimension
$d_{k,q}=\dim\mathcal{M}_{k,q}$. A basis state on the bond can
therefore be written as
\begin{equation}
    |q,\alpha,m\rangle = |N,J,\alpha,m\rangle,
    \label{eq:su2_bond_basis}
\end{equation}
where $\alpha=1,\ldots,d_{k,q}$ labels the different copies of the
spin-$J$ representation within the sector and $m=-J,\ldots,J$ is the
magnetic quantum number. The labels $(q,\alpha)$ carry the variational
information, while $m$ is fixed by symmetry and treated separately.
For a spatial fermionic orbital, the local Hilbert space decomposes
into the three $U(1)\times SU(2)$ multiplets
\begin{equation}
    (n,s)
    =
    (0,0),\qquad
    \left(1,\frac{1}{2}\right),\qquad
    (2,0).
    \label{eq:local_multiplets}
\end{equation}
In particular, $|\uparrow\rangle$ and $|\downarrow\rangle$ are the two
magnetic components of a single spin-$1/2$ multiplet rather than two
independent variational states.

Consider the tensor at site $k$. In the complete magnetic basis its
matrix elements are \cite{McCulloch2007,Singh2012SU2,WEICHSELBAUM2012,Schmoll2019SU2}
\begin{equation}
\begin{aligned}
&A^{[k]}_{(N_{k-1}J_{k-1}\alpha_{k-1}m_{k-1}),\,(n_k s_k m_{s_k}),\,(N_kJ_k\alpha_k m_k)}\\
&\quad= T^{[k]}_{(N_{k-1}J_{k-1}\alpha_{k-1}),\,(n_k s_k),\,(N_kJ_k\alpha_k)}\,
C^{J_k m_k}_{J_{k-1}m_{k-1},\,s_k m_{s_k}}
\end{aligned}
\label{eq:su2_we}
\end{equation}

where $T^{[k]}$ is the reduced MPS tensor and
$C^{J_km_k}_{J_{k-1}m_{k-1},s_km_{s_k}}$ is a Clebsch--Gordan
coefficient. This is the Wigner--Eckart decomposition of the
SU(2)-symmetric tensor.

The two factors in Eq.~\eqref{eq:su2_we} have different roles.
The reduced tensor $T^{[k]}$ contains the independent variational
parameters obtained from DMRG, whereas the CG coefficients are fixed
entirely by SU(2) representation theory.
Our synthesis procedure mirrors this factorization directly at the
circuit level.

\subsection{Reduction of the synthesized space}

The dimension of the complete spin-resolved bond
space is
\begin{equation}
    D_k
    =
    \sum_q d_{k,q}(2J+1).
    \label{eq:full_su2_dimension}
\end{equation}
In contrast, the reduced representation contains only the multiplicity
degrees of freedom,
\begin{equation}
    D_k^{\mathrm{red}}
    =
    \sum_q d_{k,q}.
    \label{eq:reduced_su2_dimension}
\end{equation}

The factors $2J+1$ are therefore removed from the space on which the
generic unitary synthesis is performed. No physical degrees of freedom
are discarded. The magnetic components are instead generated
analytically by the CG transformation introduced below.

The reduced tensor remains block sparse. \as{ A block connecting a left bond
sector $q_{L}=(N_{k-1},J_{k-1})$, a local multiplet $(n_k,s_k)$, and a
right bond sector $q_{R}=(N_k,J_k)$ can be non-zero only if}
\begin{equation}
    N_k=N_{k-1}+n_k
    \label{eq:su2_number_rule}
\end{equation}
and
\begin{equation}
    |J_{k-1}-s_k|
    \leq J_k
    \leq J_{k-1}+s_k.
    \label{eq:su2_triangle_rule}
\end{equation}
Thus, both the Abelian particle-number selection rule and the
non-Abelian angular-momentum selection rule are imposed before
compilation.

For the local multiplets of Eq.~\eqref{eq:local_multiplets}, the
spin-zero sectors satisfy
\begin{equation}
    J_k=J_{k-1},
\end{equation}
whereas a singly occupied orbital satisfies
\begin{equation}
    J_k=J_{k-1}\pm\frac{1}{2}.
    \label{eq:su2_half_rule}
\end{equation}

\subsection{Synthesis of the reduced MPS tensor}

For a fixed \as{right bond sector}
\begin{equation}
    q_R=(N_k,J_k),
\end{equation}
we collect all compatible reduced tensor blocks into an isometry
$T^{[k]}_{q_R}$. Because the original MPS is in canonical form and the CG coefficients
are orthonormal at fixed $(J_k,m_k)$, the columns of $T^{[k]}_{q_R}$ are
orthonormal, and this isometry can be completed to a unitary,
\begin{equation}
    U^{[k]}_{q_R}
    =
    \left[
        T^{[k]}_{q_R}
        \;\middle|\;
        T^{[k]}_{q_R,\perp}
    \right].
    \label{eq:su2_unitary_completion}
\end{equation}
The completion is performed independently in each symmetry sector,
giving
\begin{equation}
    U_T^{[k]}
    =
    \bigoplus_{q_R}
    U^{[k]}_{q_R}.
    \label{eq:reduced_unitary}
\end{equation}

Each $U^{[k]}_{q_R}$ is synthesized using the same Clements/QROM
procedure described in the previous section, as in the U(1) case, the blocks are synthesized in parallel and the
number of Clements layers is set by the largest block, while the QROM
of each layer loads the angles of all \as{active} blocks. The essential difference
is that the Clements decomposition now acts on the reduced
multiplicity spaces rather than on the complete spin-resolved tensor. Consequently, the dimensions of the matrices requiring generic synthesis
are reduced.

At this stage, $U_T^{[k]}$ has prepared the variational amplitudes
associated with the allowed $U(1)\times SU(2)$ coupling channels, but
the magnetic degree of freedom associated with the physical site has
not yet been explicitly emitted.

\subsection{Inverse Clebsch--Gordan transformation}

The missing magnetic structure is reconstructed using an inverse
Clebsch--Gordan transformation \cite{Sakurai2020}. For a given \as{remaining bond spin $J_{k-1}$ and local multiplet $s_k$,}
\begin{equation}
    \lambda_k = (J_{k-1}, s_k),
\end{equation}
the coupled state is transformed according to
\begin{equation}
\begin{aligned}
U_{\rm CG}^{\dagger}\,|J_k,m_k;J_{k-1},s_k\rangle
={}& \sum_{m_{k-1},\,m_{s_k}}
     C^{J_k m_k}_{J_{k-1}m_{k-1},\,s_k m_{s_k}}\\
   &\times |J_{k-1},m_{k-1}\rangle\,|s_k,m_{s_k}\rangle .
\end{aligned}
\label{eq:inverse_cg}
\end{equation}
\as{This transformation separates the physical spin of site $k$ from the spin carried by bond $k-1$.}

\begin{figure*}[!t]
\centering
\includegraphics[width=0.8\textwidth]{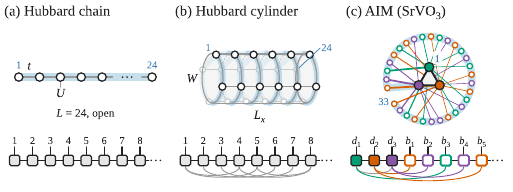}
\caption{\textbf{Benchmark systems.} (a) Open Hubbard chain of $L=24$ sites with nearest-neighbour hopping $t$ and on-site interaction $U$. (b) Hubbard cylinder of length $L_x$ with open ends and periodic width $W$. The $4\times6$ cylinder is numbered column by column and the $L_x=24$ cylinders in a snake ordering. (c) Anderson impurity model of SrVO$_3$, with three Kanamori-interacting impurity orbitals (filled circles), each coupled to ten bath orbitals of matching colour (33 sites). Numbers mark the first and last MPS sites.}
\label{fig:models}
\end{figure*}

For the empty and doubly occupied sectors, $s_k=0$, the transformation
is trivial. The only non-trivial case for a spatial orbital is
$s_k=1/2$. Writing $J=J_{k-1}$ and $M=m_k$, the two branches $J_k=J\pm\tfrac12$ with the same $M$ are mapped onto
the same two-dimensional subspace, and Eq.~\eqref{eq:inverse_cg}
reduces to a single real rotation,
\begin{equation}
\begin{aligned}
U_{\rm CG}^{\dagger}\,|J{+}\tfrac12,M\rangle
 &= \cos\theta_{J,M}\,|J,M{-}\tfrac12\rangle|{\uparrow}\rangle\\
 &\quad+ \sin\theta_{J,M}\,|J,M{+}\tfrac12\rangle|{\downarrow}\rangle,\\
U_{\rm CG}^{\dagger}\,|J{-}\tfrac12,M\rangle
 &= -\sin\theta_{J,M}\,|J,M{-}\tfrac12\rangle|{\uparrow}\rangle\\
 &\quad+ \cos\theta_{J,M}\,|J,M{+}\tfrac12\rangle|{\downarrow}\rangle,
\end{aligned}
\label{eq:CG-rotation}
\end{equation}
with $\cos\theta_{J,M}=\sqrt{(J+M+\tfrac12)/(2J+1)}$. The angle depends
only on $(J,M)$, not on $N$, $\alpha$ or the state. For $J>0$ the $2J$
values $|M|\le J-\tfrac12$ give non-trivial rotations; $M=\pm(J+\tfrac12)$
and $J=0$ only relabel the registers.

The amplitudes synthesized by
$U_T^{[k]}$ are state dependent and must be obtained from the MPS,
whereas the amplitudes appearing in $U_{\mathrm{CG}}^\dagger$ are
known analytically from the angular-momentum quantum numbers.
Because different coupling channels can occur in coherent
superposition, the CG transformation must itself be applied coherently \cite{Bacon2006}.
We implement
\begin{equation}
    \operatorname{SELECT}
    \left(U_{\mathrm{CG}}^\dagger\right)
    =
    \sum_{\lambda_k}
    |\lambda_k\rangle\langle\lambda_k|
    \otimes
    U_{\mathrm{CG}}^\dagger(\lambda_k).
    \label{eq:cg_select}
\end{equation}
No measurement of the symmetry sector is required. The bond register stores $(N,J)$, $\alpha$ and $m$ in separate
sub-registers. \as{ The CG step reads $(J,M)$ and loads $\theta_{J,M}$. The reduced channel selects $J_{k-1}$ and the emitted spin projection fixes $m_{k-1}=m_k-m_{s_k}$. Emitting $n_k$ particles leaves $N_{k-1}=N_k-n_k$ on the bond.}

The complete local SU(2)-adapted preparation operation is therefore
\begin{equation}
    V^{[k]}
    =
    \operatorname{SELECT}
    \left(U_{\mathrm{CG}}^\dagger\right)
    U_T^{[k]}.
    \label{eq:su2_local_operation}
\end{equation} 
Here $U_T^{[k]}$ contains all state-dependent variational information,
while $U_{\mathrm{CG}}^\dagger$ contains only symmetry-fixed
information. Their composition reconstructs the full tensor of
Eq.\eqref{eq:su2_we} without ever treating that
tensor as a generic dense isometry.
Applying Eq.~\eqref{eq:su2_local_operation} sequentially along the MPS
gives
\begin{equation}
    |\Psi_{\mathrm{MPS}}\rangle
    =
    \as{{V^{[1]} V^{[2]}\cdots V^{[N]}}}
    |\Psi_{\mathrm{init}}\rangle .
    \label{eq:su2_full_preparation}
\end{equation}
For a target of spin $S\neq0$ (the AIM, $S=3/2$), \as{ the initial right bond is $|N_{\rm tot},S,1,M\rangle$ in the basis of Eq.~\eqref{eq:su2_bond_basis}, and the final left bond is the vacuum.} The magnetic component $M$ is fixed by the initial state
of the $m$ register, so every component of the multiplet is prepared by
the same circuit.

\subsection{Fault-tolerant cost}
\label{sec:methods}

The cost of the SU(2)-adapted preparation separates naturally into
the synthesis of the reduced tensors and the inverse CG transformations,
\begin{equation}
    C_{SU(2)}
    =
    \sum_{k=1}^{N}
    \left[
        C\!\left(U_T^{[k]}\right)
        +
        C\!\left(
        \operatorname{SELECT}
        (U_{\mathrm{CG}}^\dagger)
        \right)
    \right].
    \label{eq:su2_cost}
\end{equation}

The first contribution is evaluated using the same QROM-assisted
Clements synthesis as for the standard and U(1) methods, but applied to the reduced blocks $U^{[k]}_{q_R}$ of Eq.~\eqref{eq:su2_unitary_completion}.
As in the U(1) case~\cite{rupprecht2026faster}, the blocks are synthesized in parallel and the
number of Clements layers is set by the largest block,
$\max_{q_R} D_{k,q_R}$, while the QROM of each layer loads the angles
\as{of the blocks active in that layer.} For the four-dimensional local Fock space of a spatial
orbital, the construction requires three applications of the
corresponding two-level synthesis procedure, giving
\begin{equation}
    C\!\left(U_T^{[k]}\right)
    =
    3\,C_{d=2}\!\left(\{D_{k,q_R}\}_{q_R};\,b\right),
    \label{eq:reduced_tensor_cost}
\end{equation}
where $\{D_{k,q_R}\}_{q_R}$ is the list of $(N,J)$ multiplicities on whichever adjacent bond has the larger maximum sector (the left bond on a tie), and
\as{$C_{d=2}$ is the per-site cost inside the sum in Eq.~\eqref{eq:total-cost}, with the QROM tables extended to the active blocks.} The U(1) estimate uses the $(N,S_z)$ sector sizes of the bond selected in the same way, and the dense estimate uses the larger expanded bond dimension.

The additional CG cost is small because the $s_k=0$ sectors require no
rotation, while the $s_k=1/2$ sectors reduce to two-dimensional
rotations with analytically known angles. Using the same $b$-bit
programmable-rotation construction as above, let
$N_{\mathrm{CG},k}$ denote the number of entries in the CG lookup table of site $k$, $2J_k+1$ for each occupied spin triple $(J_{k-1},\tfrac12,J_k)$. A QROM implementation with blocking parameter
$\Lambda_{\mathrm{CG}}$ gives
\begin{equation}
    C_{\mathrm{CG},k}
    =
    \left\lceil
        \frac{N_{\mathrm{CG},k}}
             {\Lambda_{\mathrm{CG}}}
    \right\rceil
    +
    b\Lambda_{\mathrm{CG}}
    -2.
    \label{eq:cg_cost}
\end{equation}
\as{A site with no CG angles contributes $C_{\mathrm{CG},k}=0$. 

The resulting synthesis estimate is}
\begin{equation}
    C_{\mathrm{tot}}^{SU(2)}
    =
    \sum_{k=1}^{N}
    \Big[
    3\,C_{d=2}\!\left(\{D_{k,q_R}\}_{q_R};\,b\right)
    +
    C_{\mathrm{CG},k}
    \Big].
    \label{eq:su2_total_cost}
\end{equation}
This reduction is specific to SU(2), whose multiplets carry a magnetic degeneracy absent for U(1), so only the multiplicity space needs to be synthesized.

\section{Results}
\label{sec:results}

The gate count depends on the bit-width $b$ of the rotation
angles loaded by QROM, which sets how faithfully the circuit reproduces
the MPS, and the bond dimension $\chi$, which sets the size of the
unitaries to synthesise. \as{In the benchmarks, $\chi$ is the nominal DMRG cap on the number of spin multiplets per bond. The expanded dimension $D_k$ counts all $2J+1$ magnetic components of each retained spin-$J$ multiplet [Eq.~\eqref{eq:full_su2_dimension}].} After introducing the benchmark states, we fix
$b$ from a compilation-fidelity target, report Toffoli counts, relate the
gain to the spin content of the MPS bonds, and finally compare the
compilation error with the truncation error. Throughout, the dense
(Standard) and U(1) encodings refer to the \emph{same} SU(2) MPS after
magnetic expansion, so that the three encodings prepare the same state.

\subsection{Benchmark systems}
\label{sec:benchmarks}

We consider three families of ground states that differ in the structure
of their MPS bonds (Fig.~\ref{fig:models}). One-dimensional chains with nearest-neighbour
couplings, \as{quasi-one-dimensional cylinders with longer-range couplings in the MPS ordering}, and a multi-orbital impurity model with a star
geometry and a non-singlet ground state. All Hamiltonians conserve the
particle number and are spin-rotation invariant, so that every state
carries the full $\text{U(1)}\times \text{SU(2)}$ symmetry. \as{ We use SU(2)-adapted DMRG throughout. All ground states were computed with SU(2)-adapted DMRG in Block2~\cite{block2}.}

\paragraph{Hubbard chain.}
The open one-dimensional Hubbard chain,
\begin{equation}
  H = -t \sum_{i=1}^{L-1} \sum_{\sigma}
      \left( c^\dagger_{i\sigma} c_{i+1,\sigma} + \mathrm{h.c.} \right)
      + U \sum_{i=1}^{L} n_{i\uparrow} n_{i\downarrow},
  \label{eq:hubbard}
\end{equation}
is a natural reference system for DMRG. All couplings are nearest
neighbour in the MPS ordering and, although the ground state is critical,
its entanglement grows only logarithmically with $L$. We take $L=24$ and
scan $U/t\in\{0,2,4,8\}$ at half filling ($n=1$), $U/t\in\{4,8\}$ at
$n=0.75$ and $U/t=4$ at $n=0.5$, seven singlet ground states in total,
at $\chi=800$.

\paragraph{Hubbard cylinders.}
The same Hamiltonian on a $W\times L_x$ square lattice, periodic along the
circumference ($W$ sites) and open along the length,
\begin{equation}
  H = -t \sum_{\langle \mathbf{r},\mathbf{r}'\rangle} \sum_{\sigma}
      \left( c^\dagger_{\mathbf{r}\sigma} c_{\mathbf{r}'\sigma}
      + \mathrm{h.c.} \right)
      + U \sum_{\mathbf{r}} n_{\mathbf{r}\uparrow} n_{\mathbf{r}\downarrow},
  \label{eq:hubbard_cyl}
\end{equation}
has been used to study the two-dimensional Hubbard model with
DMRG~\cite{Jiang2020}. We use two sets of cylinders. A $4\times6$ cylinder at $U/t=12$ and half filling, with the same number of sites as the chain, isolates the effect of geometry. Its sites are numbered column by column, $i=xW+y$, with every column traversed in the same direction, so that a hop along the length connects sites $W$ places apart in the MPS. To control the difficulty of the problem systematically, we use cylinders of length $L_x=24$ and widths $W=1,2,4,6$ (up to 144 sites) at $U/t=4$ and half filling. For these cylinders we use instead a snake ordering, which traverses consecutive columns in alternating directions and gave slightly better DMRG convergence. The ground
state of each cylinder is computed with DMRG up to $\chi=6400$, and the
states at smaller $\chi$ are obtained by truncating this largest-$\chi$
MPS. The cylinders with $W\le4$ converge, with a largest discarded weight
of $7\times10^{-7}$ for $W=4$ at $\chi=6400$. For $W=6$ the discarded
weight falls below $10^{-4}$ only for $\chi\ge4800$. The $W=8$ cylinder
does not converge in the accessible range of $\chi$ and is discussed in
Appendix~\ref{app:w8}.

\begin{figure}[!t]
\centering
\includegraphics[width=\columnwidth]{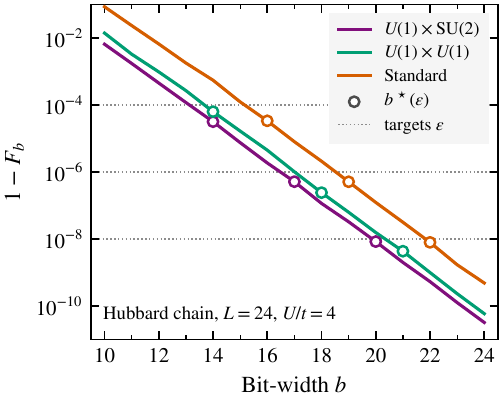}
\caption{ \textbf{Compilation infidelity $1-F_b$ versus the bit-width $b$ of the rotation angles.}
The SU(2), U(1) and Standard encodings of the same MPS at $\chi=300$ are shown for the Hubbard chain, $L=24$, $U/t=4$.
All points are exact MPS contractions. Open circles mark $b^\star(\epsilon)$, the smallest $b$
for which $1-F_b\le\epsilon$, for $\epsilon=10^{-4},10^{-6},10^{-8}$ (dotted lines).
The same behaviour is observed for all the other systems considered, including the Hubbard cylinders and the AIM.}
\label{fig:fidelity-vs-b}
\end{figure}

\begin{table*}[!t]
\centering
\caption{\as{Estimated Toffoli counts for three encodings of the same MPS.
All three use $b=21$ bits per rotation angle. At each site we choose
the QROM parameters that minimise the Toffoli count, allowing the
number of auxiliary qubits to vary.}
$D$ is the largest compilation
dimension (expanded bond, largest $(N,S_z)$ sector, largest reduced block).
Standard and U(1) refer to the \emph{same} SU(2) state after magnetic
expansion. $^\ddagger$Reduced bond saturates below $\chi=800$, at $557$ and
$746$. The half-filled $U/t=4$ chain coincides with the $W=1$ cylinder, whose
bond saturates at $\chi=315$. AIM: the $\chi=2800$ MPS and its compressions to
$\chi=300$ and $800$. Cylinders ($L_x=24$, $U/t=4$, $n=1$): largest $\chi$ per
width, discarded weight $<10^{-4}$.}
\label{tab:toffoli-all}
\small
\setlength{\tabcolsep}{4.5pt}
\begin{tabular}{l r rrr rrr rr}
\hline\hline
Model & $\chi$ & $D_{\mathrm{std}}$ & $D_{U(1)}$ & $D_{\mathrm{SU(2)}}$
 & $C_{\mathrm{std}}$ & $C_{U(1)}$ & $C_{\mathrm{SU(2)}}$
 & $C_{\mathrm{std}}/C_{\mathrm{SU(2)}}$ & $C_{U(1)}/C_{\mathrm{SU(2)}}$ \\
\hline
Chain, $U/t=0$, $n=1$ & 800 & 1931 & 205 & 122 & $4.41\times10^{7}$ & $3.40\times10^{6}$ & $1.44\times10^{6}$ & 31 & 2.4 \\
Chain, $U/t=2$, $n=1$ & 800 & 1985 & 229 & 126 & $3.44\times10^{7}$ & $2.97\times10^{6}$ & $1.18\times10^{6}$ & 29 & 2.5 \\
Chain, $U/t=8$, $n=1$ & 800$^\ddagger$ & 1513 & 180 & 105 & $2.44\times10^{7}$ & $2.29\times10^{6}$ & $8.34\times10^{5}$ & 29 & 2.7 \\
Chain, $U/t=4$, $n=0.75$ & 800 & 2025 & 235 & 133 & $4.55\times10^{7}$ & $3.90\times10^{6}$ & $1.50\times10^{6}$ & 30 & 2.6 \\
Chain, $U/t=8$, $n=0.75$ & 800 & 2092 & 248 & 134 & $5.00\times10^{7}$ & $4.24\times10^{6}$ & $1.57\times10^{6}$ & 32 & 2.7 \\
Chain, $U/t=4$, $n=0.5$ & 800$^\ddagger$ & 1846 & 231 & 120 & $2.67\times10^{7}$ & $2.55\times10^{6}$ & $1.02\times10^{6}$ & 26 & 2.5 \\
\hline
$4\times6$ cylinder, $U/t=12$ & 300 & 880 & 98 & 51 & $1.71\times10^{7}$ & $1.44\times10^{6}$ & $4.94\times10^{5}$ & 35 & 2.9 \\
 & 800 & 2477 & 250 & 124 & $7.00\times10^{7}$ & $5.08\times10^{6}$ & $1.57\times10^{6}$ & 45 & 3.2 \\
\hline
AIM (SrVO$_3$) & 300 & 1314 & 145 & 86 & $3.51\times10^{7}$ & $2.82\times10^{6}$ & $8.46\times10^{5}$ & 41 & 3.3 \\
 & 800 & 3338 & 380 & 190 & $1.40\times10^{8}$ & $9.59\times10^{6}$ & $2.79\times10^{6}$ & 50 & 3.4 \\
 & 2800 & 11557 & 1147 & 690 & $8.56\times10^{8}$ & $5.70\times10^{7}$ & $1.54\times10^{7}$ & 55 & 3.7 \\
\hline
$L_x=24$, $W=1$ & 315 & 790 & 98 & 55 & $1.00\times10^{7}$ & $9.46\times10^{5}$ & $3.71\times10^{5}$ & 27 & 2.5 \\
$L_x=24$, $W=2$ & 1200 & 3417 & 331 & 169 & $2.93\times10^{8}$ & $2.08\times10^{7}$ & $6.79\times10^{6}$ & 43 & 3.1 \\
$L_x=24$, $W=4$ & 6400 & 20010 & 1576 & 783 & $9.04\times10^{9}$ & $5.24\times10^{8}$ & $1.52\times10^{8}$ & 59 & 3.4 \\
$L_x=24$, $W=6$ & 6400 & 21107 & 1553 & 784 & $1.55\times10^{10}$ & $8.36\times10^{8}$ & $2.29\times10^{8}$ & 68 & 3.6 \\
\hline\hline
\end{tabular}
\end{table*}

\paragraph{Anderson impurity model.}
The impurity model is that of the $t_{2g}$ shell of SrVO$_3$, three
degenerate impurity orbitals $d_{a\sigma}$ ($a=1,2,3$) each hybridised
with ten bath orbitals $b_{p\sigma}$. \as{ This gives $33$ spatial orbitals, or $66$ spin orbitals. Each spatial orbital is one MPS site.} The Hamiltonian is
\begin{equation}
\begin{aligned}
  H ={}& \sum_{a,\sigma}\epsilon_d\, n_{a\sigma}
    + \sum_{p,\sigma} \epsilon_p\, b^\dagger_{p\sigma} b_{p\sigma} \\
    &+ \sum_{a,p,\sigma} \bigl( V_{ap}\, d^\dagger_{a\sigma} b_{p\sigma}
    + \mathrm{h.c.} \bigr) + H_{\rm K}.
\end{aligned}
  \label{eq:aim}
\end{equation}
The bath energies $\epsilon_p$ and hybridisations $V_{ap}$ come from an
exact-diagonalisation fit of the dynamical mean-field hybridisation
function~\cite{Georges1996}. The impurity interaction is of the rotationally invariant Kanamori form,
\begin{equation}
\begin{aligned}
  H_{\rm K} ={}& U\sum_a n_{a\uparrow}n_{a\downarrow}
  + U'\sum_{a\ne b} n_{a\uparrow}n_{b\downarrow} \\
  &+ (U'-J_H)\sum_{a<b,\sigma} n_{a\sigma}n_{b\sigma} \\
  &- J_H\sum_{a\ne b} \bigl( d^\dagger_{a\uparrow}d_{a\downarrow}
      d^\dagger_{b\downarrow}d_{b\uparrow}
    - d^\dagger_{a\uparrow}d^\dagger_{a\downarrow}
      d_{b\downarrow}d_{b\uparrow} \bigr),
\end{aligned}
  \label{eq:kanamori}
\end{equation}
with $\as{\;U=8.9143}$, $\as{\;J_H=0.6170}$ and $\as{\;U'=U-2J_{H}=7.6803}$~eV, which preserves the full
SU(2) spin symmetry. Before the DMRG, the bath block of the one-body
Hamiltonian is diagonalised, leaving the impurity orbitals untouched, and
each triply degenerate bath level is rotated so that every bath orbital
couples to a single impurity orbital. Each impurity then couples to the
bath orbitals of its own channel (star geometry \cite{Wolf2014star}), and the entanglement
concentrates on the bonds next to the impurities. We target $N=33$
electrons in the $S=3/2$ ground-state multiplet, with the impurity
orbitals first in the MPS ordering\as{, followed by the bath orbitals in order of increasing energy}, and increase the bond dimension up
to $\chi=2800$, the states at $\chi=300$ and $800$ \as{ used for the Toffoli counts} are
compressions of this MPS. The state is not fully converged at
$\chi=2800$ (discarded weight $3\times10^{-5}$, a linear extrapolation in
the discarded weight lowers the energy by $17$~meV), but our analysis
concerns the compilation of a given MPS rather than the accuracy of the
ground state.

\subsection{Choice of the angle precision}
\label{sec:res-fidelity}

The QROM loads every rotation angle with $b$ bits, i.e.\ rounded to the
nearest multiple of $2\pi/2^{b}$. \as{Rounding the reduced rotations preserves total spin when the same rotations are used for every magnetic component and the Clebsch--Gordan transformation is exact. Rounding the Clebsch--Gordan angles can mix total-spin sectors. The explicit particle-number and magnetic labels still preserve $N$ and $S_z$ in the symmetry-adapted encodings.} The compilation fidelity
$F_b=|\langle\psi|\psi_b\rangle|^2$ is obtained exactly by contracting the
two MPS, without sampling.

Figure~\ref{fig:fidelity-vs-b} compares the encodings of the same MPS for
the Hubbard chain. The infidelities decrease approximately as $4^{-b}$, as expected for independent
rounding errors of size $2^{-b}$, but they are shifted. Since $\epsilon=1-F_b$ grows with the number of rounded rotations, the encodings
that synthesise more rotations are less accurate at equal $b$. Fitting the prefactor of $4^{-b}$ over $b=14$--$22$ gives a U(1)-to-SU(2) infidelity ratio of $2.0$ and a dense-to-SU(2) ratio of $16$, using rounded CG angles. As each extra bit divides $1-F_b$ by four, this amounts
to at most one extra bit for U(1) and two for the dense encoding: SU(2) needs $b^\star=14$, $17$ and $20$ bits for $\epsilon=10^{-4}$,
$10^{-6}$ and $10^{-8}$ at $\chi=300$, against $14$, $18$, $21$ for U(1) and $16$, $19$, $22$ for the dense encoding.
The same behaviour is observed for the other systems considered, with U(1)-to-SU(2) ratios of $2.6$ for the cylinder and $2.8$ for the AIM.
With exact Clebsch--Gordan angles, the
infidelity of the reduced SU(2) circuit coincides with that of the fully
expanded one to $10^{-4}$ relative accuracy, and rounding the
Clebsch--Gordan angles to the same precision changes $1-F_b$ by less than
$0.5\%$.

For the $\chi=300$ state, SU(2) reaches $1-F_b\leq10^{-8}$ at
$b=20$ and U(1) at $b=21$. In Table~\ref{tab:toffoli-all} we compare all three
encodings at $b=21$, where the Standard encoding remains slightly above $10^{-8}$.

\subsection{Toffoli counts}
\label{sec:res-toffoli}

The non-Clifford cost is the sum over sites of the cost of synthesising
the isometry of each site, Eq.~\eqref{eq:su2_total_cost} for SU(2). A
spatial orbital has $d=4$ local states, so that in every encoding a site
costs three $d=2$ column syntheses of its block-diagonal unitary, with the
QROM of each Clements layer loading the angles \as{of its active blocks} and the QROM
parameters optimised per site (Sec.~\ref{sec:mps_prep}). The encodings differ only in their blocks. The dense encoding uses a
single block, the expanded bond, while U(1) uses the $(N,S_z)$ sectors
and SU(2) uses the reduced $(N,J)$ blocks, which in addition pays for the
SELECT of the Clebsch--Gordan angles. \as{These estimates count column synthesis and, for SU(2), the CG SELECT. The model assigns no separate cost to the boundary term $f(D_N)$, register recoding, magnetic-label updates or ancilla cleanup beyond the counted primitives.}

SU(2) is the cheapest encoding for every state in Table~\ref{tab:toffoli-all}. At $b=21$
it is $2.4$--$3.7$ times cheaper than U(1) and $26$--$68$ times cheaper
than the dense encoding, while U(1) is itself $10$--$19$ times cheaper
than the dense encoding, in line with Ref.~\cite{rupprecht2026faster}.
The two gains have different origins. Against the dense encoding, the
expanded bond is $14$--$27$ times the largest reduced block, and the
$D^{3/2}$ scaling of the column synthesis turns this into a factor of
several tens. Against U(1), the gain reflects the spin content of the
bonds. A U(1) sector $(N,S_z)$ contains one state of every multiplet
with $J\ge|S_z|$, and each multiplet appears in $2J+1$ such sectors,
whereas SU(2) keeps one block per $(N,J)$ sector.

The Toffoli count is mainly set by the bond dimension, and depends only
weakly on the angle precision $b$.
\as{At $\chi=300$, tightening the SU(2) target from $10^{-4}$ to $10^{-8}$ takes six extra bits for the chain and the $4\times6$ cylinder (14 to 20), increasing their estimated costs by about $18\%$. The AIM takes seven extra bits (13 to 20). Increasing $\chi$ from $300$ to $800$ multiplies the Table~\ref{tab:toffoli-all} costs by $3.2$ for the $4\times6$ cylinder and $3.3$ for the AIM.}

\subsection{Dependence on the bond dimension}
\label{sec:res-gain}

\begin{figure}[!tbp]
\centering
\includegraphics[width=\columnwidth]{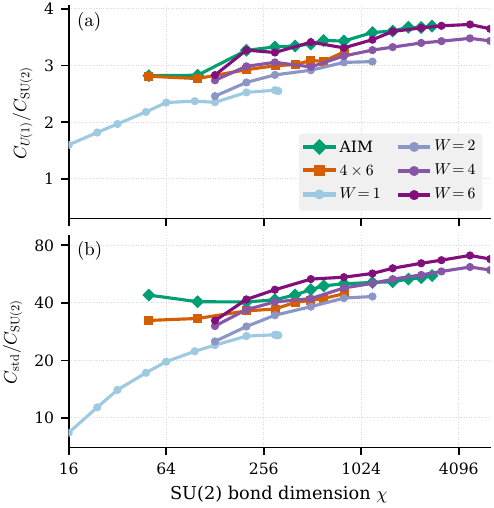}
\caption{\textbf{Toffoli-count ratios at $b=21$ versus the SU(2) bond dimension $\chi$.}
\as{(a) U(1) cost divided by SU(2) cost. (b) Dense cost divided by SU(2) cost.}
All states are truncations of a single MPS, at $\chi=800$ for the
$4\times6$ cylinder, at $\chi=2800$ for the AIM and at the largest $\chi$
of each width $W$ for the $L_x=24$ Hubbard cylinders ($U/t=4$, $n=1$).
\as{For the $4\times6$ cylinder at $\chi=300$, Table~\ref{tab:toffoli-all}
uses a different MPS from the truncation shown here.}}
\label{fig:gain-vs-chi}
\end{figure}

Figure~\ref{fig:gain-vs-chi} follows the two gains as the bond dimension
grows. All states are truncations of a single MPS, the $\chi=800$ state
for the $4\times6$ cylinder, the $\chi=2800$ state for the AIM and the
largest-$\chi$ state of each width for the $L_x=24$ cylinders. For the
AIM, the gain over U(1) increases from $2.8$ at $\chi=50$ to $3.7$ at
$\chi=2800$, and the gain over the dense encoding from about $\as{\;44}$ to
$55$. \as{For the $L_x=24$ cylinders, the gains generally increase with bond dimension and width over the sampled range.} At $\chi=300$, the gain over U(1) is
$2.6$, $2.8$, $3.1$ and $3.2$ for $W=1,2,4$ and $6$. Wider cylinders
also need a larger bond dimension to reach the same accuracy. For a \as{mean} discarded weight below $10^{-4}$ relative to the largest-$\chi$ state,
$\chi=16$ suffices for $W=1$ and $\chi=128$ for $W=2$, whereas $W=4$
requires $\chi=1200$ and $W=6$ requires $\chi=4800$. At these bond
dimensions the gain over U(1) is $1.6$, $2.5$, $3.3$ and $3.7$. At
fixed accuracy, the gain therefore grows with the difficulty of the
problem.

\begin{figure}[!tbp]
\centering
\includegraphics[width=\columnwidth]{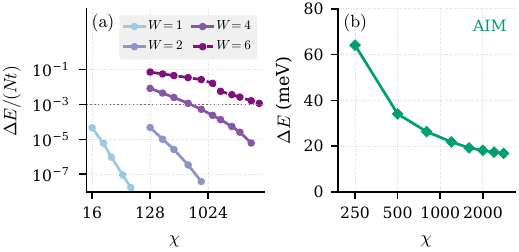}
\caption{ 
\textbf{Convergence of the DMRG energy with the bond dimension.}
(a) Energy error per site of the $L_x=24$ Hubbard cylinders ($U/t=4$,
$n=1$), relative to the largest-$\chi$ energy for $W=1$, $2$ and $4$ and
to the energy extrapolated to zero discarded weight for $W=6$ (dashed).
The dotted line marks $10^{-3}t$. (b) Energy error of the AIM relative
to the energy extrapolated to zero discarded weight. \as{We plot the last recorded energy at each $\chi$, using $-64.224173$~eV as the reference.}}
\label{fig:energy-convergence}
\end{figure}

Large bond dimensions are needed because the ground-state energy is
still converging (Fig.~\ref{fig:energy-convergence}). For $W=6$, the
DMRG energy decreases by $4.5\times10^{-3}t$ per site between
$\chi=1600$ and $\chi=6400$, and still by $5\times10^{-4}t$ per site
between $\chi=4800$ and $\chi=6400$. For the AIM, the energy decreases by
$17$~meV between $\chi=500$ and $\chi=2800$, and the extrapolation to
zero discarded weight lowers it by another $17$~meV. The DMRG states of the $W=6$ cylinder at small $\chi$, and of the $W=8$
cylinder, which in our calculations up to $\chi=4800$ remain far from
converged to the ground state, are discussed in Appendix~\ref{app:w8}.

\subsection{Compilation versus truncation error}
\label{sec:res-budget}

\as{Let $F_{\rm tot}$ be the fidelity of the prepared state with the larger-bond-dimension reference MPS, and $F_\chi$ the fidelity of the truncated MPS with that same reference. The triangle inequality for the trace distance between pure states gives $1-F_{\rm tot}\le(\sqrt{1-F_\chi}+\sqrt{1-F_b})^2$.} For the chain at $\chi=300$, $1-F_\chi=5\times10^{-12}$
with respect to $\chi=800$, the compilation dominates at every target.
For the $4\times6$ cylinder, $1-F_\chi=6\times10^{-5}$ with respect to
$\chi=800$, which already exceeds the compilation error at $b=17$. For the
AIM, with respect to $\chi=2800$, $1-F_\chi=1.5\times10^{-2}$ at
$\chi=300$ and $2.3\times10^{-3}$ at $\chi=800$. At $\chi=300$ the
\as{SU(2) compilation error} is already below $3\%$ of the truncation error for
$b=12$, and further bits are wasted. The bond dimension should therefore
be chosen from the fidelity target first, and $b$ afterwards.

\section{DISCUSSION}

Spin-rotation invariance fixes the magnetic structure of SU(2)-symmetric MPS, so the compilation presented here synthesises only
the remaining independent degrees of freedom. The reduced tensors are compiled as generic
isometries, while the magnetic structure is restored by Clebsch--Gordan
rotations whose angles are known analytically. \as{Within the cost model of
Table~\ref{tab:toffoli-all}, dense and U(1)-adapted compilation require
$26$--$68$ and $2.4$--$3.7$ times the estimated SU(2) Toffoli count,
respectively, for the same states.}

The gain has two distinct origins. First, the symmetry splits the bond into
blocks: the largest block, rather than the full bond, sets the number of Clements layers, and the QROM of each layer covers only the active blocks. Second, SU(2)
removes the magnetic degeneracy: an $(N,S_z)$ sector mixes all multiplets
with $J\ge|S_z|$, which SU(2) separates into one block per $(N,J)$ sector.
Against U(1), which already exploits the block structure, only the second
contribution remains, so that the gain grows with the spin carried by the
bonds, across models, bond dimensions and levels of convergence. Against
the dense encoding, both contributions combine.

The observed gains depend on the occupied bond-spin sectors, their multiplicities and the structural-transform overhead.
At the largest bond dimension computed, with a DMRG discarded weight below
$10^{-4}$, the gain grows from $2.4$ for the \as{$U/t=0$ chain} to $3.6$ for the six-leg
cylinder, and reaches $3.7$ for the multi-orbital impurity model. For an untruncated state, in which every bond keeps all symmetry sectors, the gains would grow faster with system size (Appendix~\ref{app:scaling}).
%
%
Increasing the bond dimension alone does not imply larger virtual spins or a larger compilation gain, and a small total spin does not bound all internal bond spins to be small. Further input families require their own compilation-cost assessment.

For the impurity problems that motivate this work, the SU(2) encoding has
a further practical advantage. Indeed, the reduced tensors do not depend on the
magnetic quantum number, so every component $M$ of the $S=3/2$ multiplet
is prepared by the same circuit, which is convenient for spin-resolved
Green's functions in DMFT~\cite{Georges1996,Jamet2025}. 
\as{In the SU(2) fidelity benchmarks, $20$--$21$ bits suffice for an infidelity of $10^{-8}$.}
The bond dimension, not the precision, controls the budget, and the two errors
should be balanced. For the impurity model the truncation error already
dominates beyond $b\simeq12$. In practice, $\chi$ should be chosen first
from the fidelity target and $b$ afterwards.

\lc{The work of Hennefarth \emph{et al.}~\cite{hennefarth2026quantumloadingchargespinadapted} also considers using SU(2) symmetries to reduce the cost compared to dense MPS state preparation, by preparing only the reduced tensors and reconstructing the magnetic degrees of freedom through Clebsch--Gordan transformations.   This work provides an alternative Clements-based construction and provides a direct comparison against U(1)-adapted compilation using the Clements-based framework of Ref.~\cite{rupprecht2026faster}. By using the same compilation framework, we highlight the additional benefits provided by using non-Abelian symmetries.}

The method has limitations. It requires exact spin-rotation invariance. A magnetic field along $z$ reduces the symmetry to $\mathrm{U}(1)_N\times\mathrm{U}(1)_{S_z}$, for which the compilation of Ref.~\cite{rupprecht2026faster} remains the reference, whereas generic spin--orbit coupling also breaks $S_z$ conservation. It also prepares a state of definite total spin, which must be
chosen before the DMRG. When the spin of the ground state is not known,
several spin sectors must be computed and compared, and for systems with
nearly degenerate spin states the method prepares one of them rather than
a superposition. Our counts are analytical, following Ref.~\cite{berry2025rapid}, and the
column synthesis uses the same factor $d-1=3$ for all three encodings.

Several extensions are natural. The real-valued synthesis of
Ref.~\cite{rupprecht2026faster} acts on the reduced tensors as well and can be
combined with the present scheme. \as{The same construction applies when the symmetry acts on each MPS site and the tensors respect that action. The compilation gain then depends on the cost of the corresponding change of basis.} Examples include point-group symmetries of
molecules, translations around a cylinder \as{when each ring is grouped into one MPS site}, and the pseudospin SU(2) of
the half-filled Hubbard model. Impurity models often have larger
symmetry groups, such as the orbital SO(3) and spin SU(2) symmetry of
the three-orbital Kanamori interaction~\cite{Horvat2016} or the $SU(3)$
symmetry of the multichannel Anderson model~\cite{Moca2012}, which are
already exploited by numerical renormalization group solvers. Extending
the compilation to these groups could further reduce the cost of
preparing impurity ground states. 


\begin{acknowledgments}
%
L.P.L. acknowledges support from the UK Department for Science Innovation and Technology through the UK National Quantum Technologies Programme. L.P.L. thanks Abhishek Agarwal for useful comments on the manuscript.

\end{acknowledgments}


\bibliography{Biblio}

\onecolumngrid

\appendix

\section{Toffoli cost of the column synthesis}
\label{app:complexity}

We count the non-Clifford gates needed to prepare one site tensor,
i.e.\ to implement the unitary $U_{T_i}$ of dimension $N_{\rm un}$ that
completes the isometry $T_i$. For a local dimension $d=2$,
\as{$D_i=\max(\chi_{i-1},\chi_i)$ is the active bond dimension. The padded unitary has dimension $N_{\rm un}=2^n$, with $n=\lceil\log_2(2D_i)\rceil$.} The synthesis of Sec.~\ref{sec:mps_prep}
proceeds in three steps, $b$ denotes the bit-width of the rotation
angles stored in the QROM, and $\Lambda,\Lambda'$ are the
space--time trade-off parameters of the QROM of
Ref.~\cite{Low2024LKS}.

\paragraph{Step 1: block-diagonalising unitary.}

If $f(D)$ denotes
the cost of a generic $D$-dimensional unitary synthesis, this
contributes $f(D_i)$. The contribution does not accumulate, however
\as{ for $i<N$ the product $W_i\,{\rm diag}(U_1,U_2)_{i+1}$, where $W_i$ is the block-diagonalising unitary,} can be
absorbed into the previous layer, so only the first layer pays the
full price $\as{{f(D_N)}}$.

\paragraph{Step 2: QROM-controlled rotation.}
The remaining qubits receive a rotation controlled on the register.
Loading $D_i$ angles of $b$ bits with a QROM costs
$\lceil D_i/\Lambda\rceil+b(\Lambda-1)$ Toffolis~\cite{Low2024LKS}, and
the controlled rotation itself costs $b-2$, hence
\begin{equation}
  C_{\rm rot}(D_i)=\Big\lceil\frac{D_i}{\Lambda}\Big\rceil+b\Lambda-2 .
  \label{eq:app-rot}
\end{equation}

\paragraph{Step 3: block-diagonal unitary.}
It remains to implement
$\begin{pmatrix}U_1&0\\0&U_2\end{pmatrix}$. \as{The estimate counts $D_i$ phasing layers.} The QROM is addressed by the pair index together with the qubit that selects $U_1$ or $U_2$, and the increment/decrement
operations act on the pair index and the target qubit, so each uses $n-1$ qubits. Applying the general
unitary-synthesis count to each layer gives
\begin{align}
  C_{\rm bd}(D_i)
  ={}&D_i\Big(\Big\lceil\frac{D_i}{\Lambda}\Big\rceil+2\Lambda b-5\Big)
  &&\text{($D_i$ phasing layers)}
  \label{eq:app-layers}\\
  &+(n-3)(D_i-1)
  &&\text{(INCR/DECR on $n-1$ qubits)}
  \label{eq:app-incr}\\
  &+\Big\lceil\frac{D_i}{\Lambda}\Big\rceil+2\Lambda b
    +\Big\lceil\frac{2D_i}{\Lambda'}\Big\rceil+\Lambda'-6
  &&\text{(phase shifts).}
  \label{eq:app-phase}
\end{align}

\paragraph{Total.}
Summing over the $N$ sites of the MPS,
\begin{equation}
  C_{\rm tot}=\as{{f(D_N)}}+\sum_{i=1}^{N}\big[C_{\rm rot}(D_i)+C_{\rm bd}(D_i)\big],
  \label{eq:app-total}
\end{equation}
with $\Lambda$ and $\Lambda'$ optimised independently at every site
over the integers. The dominant term is Eq.~\eqref{eq:app-layers},
$D_i(\lceil D_i/\Lambda\rceil+2\Lambda b)$, minimised by
$\Lambda\simeq\sqrt{D_i/2b}$, which gives
$\simeq2\sqrt2\,D_i^{3/2}b^{1/2}$ per site. The total Toffoli count
therefore scales as
\begin{equation}
  C_{\rm tot}=O\!\big(N\,\chi^{3/2}\,b^{1/2}\big),
  \label{eq:app-scaling}
\end{equation}
polynomial in the bond dimension and only square-root in the angle
precision. \as{Equation~\eqref{eq:app-total} describes a single block. For the block-diagonal synthesis in Table~\ref{tab:toffoli-all}, the largest block sets the number of Clements layers. In each layer the QROM loads the angles for all blocks that are still active. The SU(2) estimate also includes the SELECT of the Clebsch--Gordan angles.}

\section{Non-converged $W=8$ cylinder}
\label{app:w8}

In the main text, the states of the $L_x=24$ cylinders at each bond
dimension are truncations of the largest-$\chi$ DMRG state of each
width. Here we consider instead the DMRG states obtained directly at
each $\chi$. For $W=1$, $2$ and $4$, these states converge smoothly to
the largest-$\chi$ state. For $W=6$, the states at $\chi\le1200$ have a
fidelity below $2\times10^{-3}$ with the $\chi=6400$ state, which rises
to $0.80$ at $\chi=1600$ and $0.99$ at $\chi=4800$. \as{The $W=6$ states first show substantial overlap with the largest-$\chi$ reference at $\chi=1600$.} The $W=8$ cylinder does not converge
in the range of $\chi$ that we have considered. Its fidelity between the $\chi=3200$
and $\chi=4800$ states is only $0.32$, and its energy per site still
decreases by $5\times10^{-3}t$ over this range.

\begin{figure}[!t]
\centering
\includegraphics[width=\columnwidth]{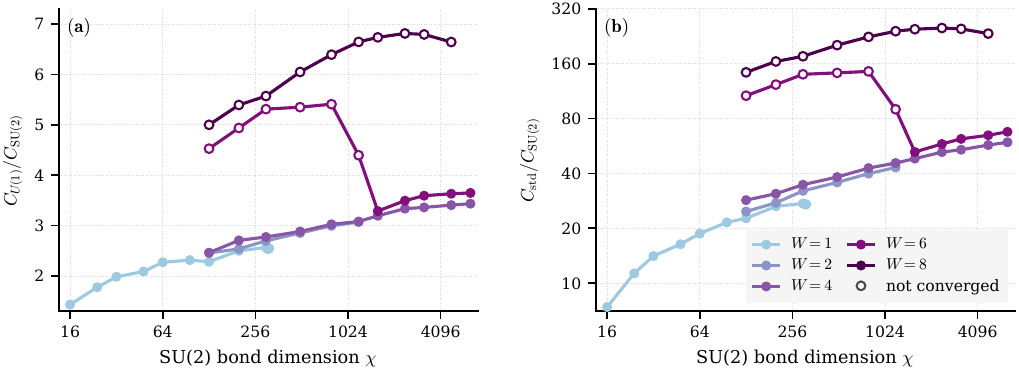}
\caption{ \textbf{Gain of the DMRG states of the $L_x=24$ Hubbard cylinders ($U/t=4$, $n=1$) versus the bond dimension.}
\as{(a) U(1) cost divided by SU(2) cost. (b) Dense cost divided by SU(2) cost.}
Each point is a separate DMRG state at bond dimension $\chi$. Open
symbols \as{mark states with fidelity below $0.5$ relative to the largest-$\chi$ state, and all $W=8$ states.}}
\label{fig:dmrg-gain}
\end{figure}

\as{The low-overlap $W=6$ states have a larger fraction of high-spin bond multiplets than truncations of the largest-$\chi$ reference at the same $\chi$. Their U(1)-to-SU(2) cost ratio ranges from $4.4$ to $5.4$ through $\chi=1200$, and falls to $3.3$ at $\chi=1600$ (Fig.~\ref{fig:dmrg-gain}). For $W=8$, the ratios at the last sampled point ($\chi=4800$) are $6.6$ for U(1) and $234$ for the dense encoding.}

\section{Compilation Method}
\label{app:compil}

\begin{figure}[!t]
\begin{minipage}[t]{\dimexpr(\textwidth-\columnsep)/2\relax}
\vspace{0pt}\centering
\begin{tikzpicture}[
    x=0.85cm,
    y=0.85cm,
    line/.style={draw, line width=0.7pt},
    gate/.style={
        draw,
        minimum width=0.65cm,
        minimum height=0.65cm,
        inner sep=1pt,
        font=\small
    },
    phase/.style={
        draw,
        minimum width=0.45cm,
        minimum height=0.45cm,
        inner sep=0pt,
        font=\scriptsize
    }
]


\begin{scope}[xshift=0cm,yshift=0cm]

\node at (-0.5,4.8) {(a)};

\foreach \y in {4.6,3.9,3.2,2.5,1.8,1.1,0.4,-0.3}
    \draw[line] (0,\y) -- (0.45,\y);


\foreach \y in {4.6,3.2,1.8,0.4}{
    \draw[line]
        (0.45,\y)
        -- (0.95,\y-0.35)
        -- (1.45,\y);

    \draw[line]
        (0.45,\y-0.7)
        -- (0.95,\y-0.35)
        -- (1.45,\y-0.7);
}


\draw[line] (1.45,4.6) -- (2.15,4.6);
\draw[line] (1.45,-0.3) -- (2.15,-0.3);

\foreach \y in {3.9,2.5,1.1}{
    \draw[line]
        (1.45,\y)
        -- (1.80,\y-0.35)
        -- (2.15,\y);

    \draw[line]
        (1.45,\y-0.7)
        -- (1.80,\y-0.35)
        -- (2.15,\y-0.7);
}


\foreach \y in {4.6,3.2,1.8,0.4}{
    \draw[line]
        (2.15,\y)
        -- (2.65,\y-0.35)
        -- (3.15,\y);

    \draw[line]
        (2.15,\y-0.7)
        -- (2.65,\y-0.35)
        -- (3.15,\y-0.7);
}


\draw[line] (3.15,4.6) -- (3.85,4.6);
\draw[line] (3.15,-0.3) -- (3.85,-0.3);

\foreach \y in {3.9,2.5,1.1}{
    \draw[line]
        (3.15,\y)
        -- (3.50,\y-0.35)
        -- (3.85,\y);

    \draw[line]
        (3.15,\y-0.7)
        -- (3.50,\y-0.35)
        -- (3.85,\y-0.7);
}


\foreach \y in {4.6,3.2,1.8,0.4}{
    \draw[line]
        (3.85,\y)
        -- (4.35,\y-0.35)
        -- (4.85,\y);

    \draw[line]
        (3.85,\y-0.7)
        -- (4.35,\y-0.35)
        -- (4.85,\y-0.7);
}


\draw[line] (4.85,4.6) -- (5.55,4.6);
\draw[line] (4.85,-0.3) -- (5.55,-0.3);

\foreach \y in {3.9,2.5,1.1}{
    \draw[line]
        (4.85,\y)
        -- (5.20,\y-0.35)
        -- (5.55,\y);

    \draw[line]
        (4.85,\y-0.7)
        -- (5.20,\y-0.35)
        -- (5.55,\y-0.7);
}


\foreach \y in {4.6,3.2,1.8,0.4}{
    \draw[line]
        (5.55,\y)
        -- (6.05,\y-0.35)
        -- (6.55,\y);

    \draw[line]
        (5.55,\y-0.7)
        -- (6.05,\y-0.35)
        -- (6.55,\y-0.7);
}

\foreach \y in {4.6,3.9,3.2,2.5,1.8,1.1,0.4,-0.3}
    \draw[line] (6.55,\y) -- (7.0,\y);

\end{scope}


\begin{scope}[xshift=0cm,yshift=-2.4cm]

\node at (-0.5,1.2) {(b)};


\draw[line]
    (0,1.2)
    -- (0.55,1.2)
    -- (1.05,0.7)
    -- (1.55,1.2)
    -- (2.05,1.2);

\draw[line]
    (0,0.2)
    -- (0.55,0.2)
    -- (1.05,0.7)
    -- (1.55,0.2)
    -- (2.05,0.2);

\node at (2.55,0.7) {$=$};


\draw[line]
    (3.0,1.2)
    -- (3.45,1.2)
    -- (3.85,0.7)
    -- (4.25,1.2);

\draw[line]
    (3.0,0.2)
    -- (3.45,0.2)
    -- (3.85,0.7)
    -- (4.25,0.2);

\node[phase] at (3.45,1.65) {$\phi-\pi/2$};
\draw[line] (3.45,1.425) -- (3.45,1.2);

\draw[line] (4.25,1.2) -- (4.8,1.2);
\draw[line] (4.25,0.2) -- (4.8,0.2);

\draw[line]
    (4.8,1.2)
    -- (5.25,0.7)
    -- (5.7,1.2);

\draw[line]
    (4.8,0.2)
    -- (5.25,0.7)
    -- (5.7,0.2);

\node[phase] at (4.8,1.65) {$2\theta$};
\draw[line] (4.8,1.425) -- (4.8,1.2);

\draw[line] (5.7,1.2) -- (6.55,1.2);
\draw[line] (5.7,0.2) -- (6.55,0.2);

\node[phase] at (6.15,1.65) {$\varphi_0$};
\draw[line] (6.15,1.425) -- (6.15,1.2);

\node[phase] at (6.15,-0.25) {$\varphi_1$};
\draw[line] (6.15,-0.025) -- (6.15,0.2);

\end{scope}


\end{tikzpicture}

\smallskip
\begin{tabular}{@{}c@{\hspace{0.5em}}c@{}}
(c) &
\begin{quantikz}[column sep=0.2cm]
 & \gate{R(\phi-\pi/2)} & \gate{H} & \gate{R(2\theta)} & \gate{H} & \gate{\mathrm{diag}(e^{i\varphi_0},e^{i\varphi_1})} &
\end{quantikz}
\end{tabular}
\caption{Decomposition of a unitary transformation into an equivalent quantum
circuit.
(a) Rectangular mesh decomposition into a sequence of two-mode transformations.
(b) Each tunable two-mode transformation is decomposed into fixed $50{:}50$
beam splitters, here Hadamard gates $H$, and phase shifts.
(c) Equivalent quantum-circuit representation, where the $50{:}50$ beam
splitters are mapped to Hadamard gates and the phase shifts to single-qubit
rotation gates $R(\alpha)=\mathrm{diag}(e^{i\alpha},1)$, and the phases pushed to the end of the decomposition.}

\label{fig:mesh}
\end{minipage}\hfill%
\begin{minipage}[t]{\dimexpr(\textwidth-\columnsep)/2\relax}
\vspace{0pt}\centering
\resizebox{\linewidth}{!}{%
\begin{tabular}{@{}l@{}}
(a)\\[2pt]
\begin{quantikz}[row sep={0.75cm,between origins}, column sep=0.15cm]
\lstick{address} & \qwbundle{n-1} &[0.55cm] \gate[3]{\mathrm{QROM}} & & & & & \gate[3]{\mathrm{QROM}^{\dagger}} & \\
\lstick{$\phi$ register} & \qwbundle{b} & & \ctrl{2} & & & & & \\
\lstick{$\theta$ register} & \qwbundle{b} & & & & \ctrl{1} & & & \\
\lstick{target} & & & \gate{R(\phi_j-\pi/2)} & \gate{H} & \gate{R(2\theta_j)} & \gate{H} & &
\end{quantikz}\\[14pt]
(b)\\[2pt]
\begin{quantikz}[row sep={0.75cm,between origins}, column sep=0.15cm]
\lstick{address} & \qwbundle{n-1} &[0.55cm] \gate[2]{\text{Layer }1} & \gate[2]{\mathrm{INCR}} & \gate[2]{\text{Layer }2} & \gate[2]{\mathrm{DECR}} & \gate[2]{\text{Layer }3} & \gate[2]{\text{Phasing}} & \\
\lstick{target} & & & & & & & &
\end{quantikz}
\end{tabular}%
}
\caption{Compilation of the rectangular interferometer mesh into a quantum circuit.
(a) One Clements layer. The QROM loads the angles $\phi_j$ and $\theta_j$ of the
two-mode transformation addressed by $j$ into $b$-bit registers, which control the
rotations on the target qubit. The registers are then uncomputed.
(b) Full circuit for a mesh of three layers followed by the final phasing layer.
The address and target together are incremented or decremented between layers to switch between even and odd pairs. Writing $x=2j+t$ for the mode with target bit $t$, adding one maps the odd pairs $(2j+1,2j+2)$ onto even pairs.}
\label{fig:clements_quantum_circuit}
\end{minipage}
\end{figure}

\as{Numbering the $\ell$ layers in the order in which they act gives
\begin{equation}
    U = \quad \Phi L_\ell \cdots L_2 L_1 ,
    \label{eq:clements_decomposition}
\end{equation}}
where each $\as{\;L_j}$ is a layer of mutually commuting two-level
transformations acting on disjoint pairs of modes, and $\Phi$ is a
final diagonal phase matrix. The layers alternate between the pairs
$(0,1),(2,3),\ldots$ and $(1,2),(3,4),\ldots,$
forming the rectangular Clements mesh.

Each two-level transformation is completely specified by a small
number of rotation angles. In the convention used here, it can be
written, up to phase conventions, as
\begin{equation}
\begin{pmatrix}
e^{i\phi}\cos\theta & -\sin\theta \\
e^{i\phi}\sin\theta & \cos\theta
\end{pmatrix}
=\mathrm{diag}(e^{i\varphi_0},e^{i\varphi_1})\,HR(2\theta)HR(\phi-\pi/2),
\label{eq:clements_block}
\end{equation}
with $\varphi_0=\pi/2-\theta$ and $\varphi_1=-\theta$. This corresponds to the circuit of Fig.~\ref{fig:mesh}(c).
Consequently, a dense $N\times N$ unitary requires $\mathcal{O}(N^2)$ two-level
transformations arranged in $\mathcal{O}(N)$ layers.

The rotation angles entering the Clements decomposition are computed
classically and encoded with $b$ bits of precision. Rather than
synthesizing each rotation independently, Ref.~\cite{berry2025rapid}
uses quantum read-only memory (QROM) \cite{Babbush2018} to coherently load the angles
associated with the state of an address register: $
    |j\rangle |0\rangle_{\theta}|0\rangle_{\phi}
    \xrightarrow{\text{QROM}}
    |j\rangle |\theta_j\rangle|\phi_j\rangle$.\\
    \\
The loaded angles are subsequently used to implement the corresponding rotations, via phase gradient rotation \cite{kitaev2002classical}. This construction allows all two-level
transformations belonging to a Clements layer to be addressed through
the same circuit structure Fig.\ref{fig:clements_quantum_circuit}, while the required rotation parameters are supplied through QROM. Alternating layers are obtained by shifting the addressing of the neighboring modes, so that the complete Clements mesh can be implemented sequentially.

\section{Scaling of the gain with system size}\label{app:scaling}

\begin{figure}[h]
\centering
\includegraphics[width=0.5\columnwidth]{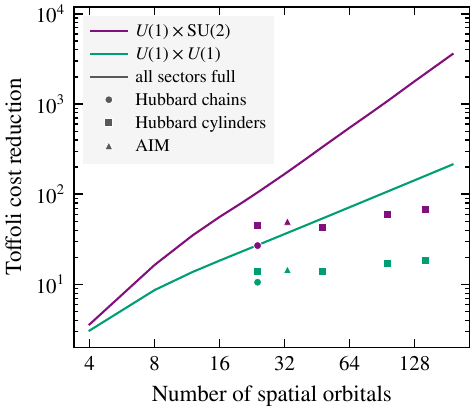}
\caption{\textbf{Toffoli-count reduction relative to the dense encoding versus the number of spatial orbitals $L$.} Lines show a half-filled singlet with maximal multiplicity in every $(N,J)$ bond sector, for the U(1) (green) and SU(2) (purple) encodings. All counts use $b=21$.}
\label{fig:scaling}
\end{figure}

Figure~\ref{fig:scaling} compares the Toffoli-count reductions of the benchmarks with those of a half-filled singlet with maximal multiplicity in every $(N,J)$ bond sector. Applying the active-block QROM rule of Eq.~\eqref{eq:reduced_tensor_cost} to the layer cost of Eq.~\eqref{eq:app-layers} and optimising the QROM blocking gives a leading site cost proportional to $\sqrt{D_{\max}\sum_k D_k^2}$ at fixed angle precision, where $D_k$ are the block sizes. At the middle cut this gives asymptotic reductions relative to the dense encoding of $\pi L/(2\sqrt{2})$ for $\mathrm{U}(1)\times\mathrm{U}(1)$ and $0.36\,L^{7/4}$ for $\mathrm{U}(1)\times\mathrm{SU}(2)$, where $L$ is the number of spatial orbitals. The DMRG ground states lie below these curves, since a finite $\chi$ limits the sector multiplicities, and the doped chains and the impurity model lie in $(N,S)$ sectors different from the half-filled singlet.

\end{document}